\documentclass[prl,twocolumn,aps,preprintnumbers,10pt]{revtex4-1}

\usepackage{graphicx}
\usepackage{amsmath}
\usepackage{dsfont}
\usepackage{amsfonts}
\usepackage{filecontents}
\usepackage{slashed}
\usepackage{microtype}
\usepackage[macrosonly]{chet}
\usepackage{xcolor}
\definecolor{darkblue}{rgb}{0.1,0.1,.7}
\definecolor{darkgreen}{rgb}{0,.5,0}
\usepackage[pdftex,breaklinks,colorlinks,linkcolor=darkblue,
citecolor=darkgreen]{hyperref}

\newcommand{\cO}{{\cal O}}

\DeclareMathOperator{\Tr}{Tr}

\renewcommand{\a}{\alpha}
\renewcommand{\b}{\beta}

\newcommand{\m}{\mu}
\newcommand{\n}{\nu}

\def\be{\begin{equation}}
\def\ee{\end{equation}}
\def\bea{\begin{eqnarray}}
\def\eea{\end{eqnarray}}
\def\ba{\begin{array}}
\def\ea{\end{array}}
\def\bi{\begin{itemize}}
\def\ei{\end{itemize}}

\def\Tr{{\rm Tr}}

\newcommand{\beq}{\begin{equation}}
\newcommand{\eeq}{\end{equation}}
\newcommand{\beqn}{\begin{eqnarray}}
\newcommand{\eeqn}{\end{eqnarray}}
\newcommand{\bga}{\begin{align}}

\def\dalemb#1#2{{\vbox{\hrule height .#2pt
\hbox{\vrule width.#2pt height#1pt \kern#1pt
\vrule width.#2pt}
\hrule height.#2pt}}}

\def\a{\alpha}
\def\b{\beta}

\def\bz{{\overline{z}}}

 \def\IZ{\relax\ifmmode\mathchoice
 {\hbox{\cmss Z\kern-.4em Z}}{\hbox{\cmss Z\kern-.4em Z}}
 {\lower.9pt\hbox{\cmsss Z\kern-.4em Z}}
 {\lower1.2pt\hbox{\cmsss Z\kern-.4em Z}}\else{\cmss Z\kern-.4em Z}\fi}
 \def\IB{\relax{\rm I\kern-.18em B}}
 \def\IC{{\relax\hbox{$\inbar\kern-.3em{\rm C}$}}}
 \def\Ic{{\relax\hbox{$\inbar\kern-.22em{\rm c}$}}}
 \def\ID{\relax{\rm I\kern-.18em D}}
 \def\IE{\relax{\rm I\kern-.18em E}}
 \def\IF{\relax{\rm I\kern-.18em F}}
 \def\IG{\relax\hbox{$\inbar\kern-.3em{\rm G}$}}
 \def\IGa{\relax\hbox{${\rm I}\kern-.18em\Gamma$}}
 \def\IH{\relax{\rm I\kern-.18em H}}
 \def\II{\relax{\rm I\kern-.18em I}}
 \def\IK{\relax{\rm I\kern-.18em K}}
 \def\IP{\relax{\rm I\kern-.18em P}}

\def\Tr{{\rm Tr}}
 \font\cmss=cmss10 \font\cmsss=cmss10 at 7pt
 \def\IR{\relax{\rm I\kern-.18em R}}

\def\cO{{\cal O}}

\def\a{\alpha}
\def\b{\beta}

\def\m{\mu}
\def\n{\nu}

\def\cQ{{\cal Q}}

\begin{document}

\preprint{CPHT-RR069.112023}
\title{Conformal graphs as twisted partition functions}
\author{Manthos Karydas$^a$}
\author{Songyuan Li $^b$}
\author{Anastasios C. Petkou $^b$}
\author{and Matthieu Vilatte $^{b,c}$}
\affiliation{$^a$ Illinois Center for Advanced Studies of the Universe \& Department of Physics, University of Illinois, 1110 West Green St., Urbana IL 61801, USA}
\affiliation{$^b$ Division of Theoretical Physics, School of Physics,
  Aristotle University of Thessaloniki, 54124 Thessaloniki, Greece}
\affiliation{$^c$ Centre de Physique Th\'eorique -- CPHT,
        \'Ecole polytechnique, CNRS -- Unit\'e Mixte de Recherche UMR 7644,
        Institut Polytechnique de Paris, 91120 Palaiseau Cedex, France}
\date{\today}
\begin{abstract}
We show that a class of $L$-loop conformal ladder graphs are intimately related to twisted partition functions of free massive complex scalars in $d=2L+1$ dimensions. The graphs arise as four-point functions in certain two- and four-dimensional conformal fishnet models.  The twisted thermal two-point function of the scalars becomes a generator of conformal ladder graphs for all loops. We argue that this  correspondence is seeded by a system of two decoupled harmonic oscillators twisted by an imaginary chemical potential. We find a number of algebraic and differential relations among the conformal graphs which mirror the underlying free dynamics.  
\end{abstract}

\maketitle

\emph{Introduction and summary.}---In \cite{Petkou:2021zhg}, one of the authors observed that the logarithm of the partition function $Z_L$ of a free massive complex scalar $\phi(x)$, twisted by the global $U(1)$ charge along the thermal circle in $d=2L+1$ dimensions, is given in terms of a class of single-valued polylogarithms. The latter functions are ubiquitous in multiloop QFT calculations (see e.g. \cite{DelDuca:2022skz} for a recent review), and their intriguing mathematical properties have been discussed  in a number of works \cite{BROWN2004527,Schnetz:2013hqa}.  The twisting parameter $\mu$ corresponds to an imaginary chemical potential for the abelian ``charge" operator $\cQ=\phi^\dagger\overleftrightarrow{{\cal D}_\tau}\phi$, with ${\cal D}_\tau=\partial_\tau-i\mu$, which together with $\cO=|\phi|^2$ can be viewed as integrable relevant deformations of the massless free theory. From $\ln Z_L$ we can calculate the thermal one-point functions  $\langle\cO\rangle_L$ and $\langle\cQ\rangle_L$ respectively, and it was shown in \cite{Petkou:2021zhg} that $\langle\cQ\rangle_L $ is essentially given by the $L$-loop Davydychev-Usyukina conformal ladder graph \cite{Usyukina:1992jd,Usyukina:1993ch}. 

We show here that $\ln Z_L$ itself is also given by an $L$-loop conformal ladder graph which evaluates a certain four-point function of the singular two-dimensional conformal fishnet model of Kazakov and Olivucci \cite{Kazakov:2018qbr}. Consequently, the differential equations satisfied by $\langle\cO\rangle_L$ and $\langle\cQ\rangle_L$ presented in \cite{Petkou:2021zhg} become differential relations among four-point ladder graphs of conformal fishnet models in two- and four-dimensions. The observations above prompt us to consider the twisted thermal two-point function $\langle\phi^\dagger(x)\phi(0)\rangle_L$. When $m=\mu=0$ this is expanded in thermal conformal blocks with constant coefficients corresponding to the thermal one-point functions of conformal quasiprimary operators with definite dimension and spin (see (\ref{phiphi}) later on).  We show that for nonzero values of $m$ and $\mu$ the above two-point function can also be expanded in terms of thermal conformal blocks, but with coefficients now given by single-valued polylogarithms. The latter are recursively related to linear  combinations of $\langle\cO\rangle_L$ and $\langle\cQ\rangle_L$, and hence of conformal ladder graphs. In other words $\langle\phi^\dagger(x)\phi(0)\rangle_L$ is a generating function of all-loop conformal ladder graphs. 
Some implications of our results and a number of future directions are discussed. 


\emph{From relativistic Bose gases to single-valued polylogarithms}---
 We firstly rederive the results in \cite{Petkou:2021zhg} from a new perspective. Consider the twisted partition function of two decoupled harmonic oscillators with unit mass and common frequency $m$
\be
\label{Z0def}
Z_0=\Tr_{{\cal H}_{1,2}} \left[e^{-\beta (H_0+m^2\cO)}e^{-i\beta\mu\cQ}\right]\,.
\ee
This can be  viewed as a deformation of the free Hamiltonian  $H_0=(\hat{p}^2_1+\hat{p}^2_2)/2$  by the operators $\cO=\frac{1}{2}(\hat{x}^2_1+ \hat{x}^2_2)$ and $\cQ=\hat{p}_{2}\hat{x}_1-\hat{p}_{1}\hat{x}_2$.
\footnote{We set $\hbar=1$ and define the (creation) annihilation operators $\hat{a}_{1}=\frac{1}{2\sqrt{m}}(m(\hat{x}_{1}-i\hat{x}_{2})+ (\hat{p}_{2}+i\hat{p}_{1}))\,,\,\,\hat{a}_{2}=\frac{1}{2\sqrt{m}}(m(\hat{x}_{1}+i\hat{x}_{2})- (\hat{p}_{2}-i \hat{p}_{1}))\,,\,\,\,[\hat{x}_i,\hat{p}_j]=i\delta_{ij}\,,\,[\hat{a}_i,\hat{a}_j^\dagger]=\delta_{ij}\,,\,i=1,2$.

The trace in (\ref{Z0def}) is taken over the tensor product Hilbert space ${\cal H}_{1,2}\approx \{|n_1\rangle\otimes|n_2\rangle\}$, $n_1,n_2=0,1,2,..$.} 
The twisting parameter $\mu$ acts effectively as an imaginary chemical potential for $\cQ$. $Z_0$ is the grand canonical partition function. Using the complex variable $z=e^{-\beta m-i\beta\mu}$ one finds
\begin{equation}
\label{Z0}
    \ln Z_0=\int_0^{z}\frac{dz'}{1-z'}+\int_0^{\bz}\frac{dz'}{1-z'}-\int_{|z|}^1\frac{dz'}{z'}\,.
\end{equation}
From (\ref{Z0}) we can construct the logarithm of the partition function of a free charged scalar field in $d$-dimensions with mass $m$ and twisting parameter $\mu$ \footnote{\label{rho}For a field theory defined on the Euclidean thermal geometry $S_\beta^1\times\mathbb{R}^{d-1}$ the parameter $\mu$ could  be either considered as a twisting $\phi(\tau+\beta,\vec{x})=e^{i\beta\mu}\phi(\tau,\vec{x})$, or as an imaginary chemical potential, or as the $\tau$ component of a real background gauge potential.} as that of a $d=2L+1$-dimensional relativistic thermal gas (see Appendix B for details)
\begin{equation}
\label{Zd}
    \ln Z_L=\int d\omega \,\rho_L(\omega;m)\ln Z_0\,.
\end{equation}
The calculations are considerably simpler for integer $L$ ($d$ odd) to which we restrict from now on. 

 After some straightforward manipulations (\ref{Zd})  can be brought into the form of an iterated integral for $L>1$ as
\begin{equation}
\label{ZL}
    \ln Z_L=(-2\alpha^2)^L\prod_{i=0}^{L-1}\left[\int_0^{w_{i+1}}\frac{dw_i}{w_i}\ln w_i\right]\ln Z_0\,,
\end{equation}
where $\ln Z_0$ is taken to be a function of $z_0,\bar{z}_0$ with \mbox{$z_0=w_0e^{-i\beta\mu}$}, $w_{L}=|z|$ and the integrals are performed in the order $w_0\mapsto w_1..\mapsto w_{L}$. Here $\alpha^2=\ell^2/4\pi\beta^2$ is a dimensionless parameter. By virtue of (\ref{Z0}) 
we see that (\ref{ZL}) coincides with the class of iterated integrals that give rise to single-valued polylogarithms \cite{Schnetz:2013hqa}. We  obtain \footnote{It can be shown that the above procedure yields the same results with the corresponding path integral calculation The infinite result coming from the zero-point energies of the harmonic oscillators can be consistently regularized by subtracting the zero temperature partition function for $m=\mu=0$.}
\begin{align}
\label{ZLresult}
    \ln Z_L&=\a^{2L}\frac{(-1)^LL!}{2(2L+1)!}(2\log|z|)^{2L+1}\nonumber \\
    &\hspace{-.8cm}+\a^{2L}\sum_{n=0}^{L} \frac{ (2L-n)! (-2\log |z|)^n}{(L-n)!n!} 2\Re[Li_{2L+1-n}(z)]\,,\\
\label{QLresult}
    &\hspace{-1.1cm}\langle Q\rangle_L = \alpha^{2L}\sum_{n=0}^{L} \frac{ (2L-n)! (-2\log |z|)^n}{(L-n)!n!} 2i\Im[Li_{2L-n}(z)].
\end{align}
The formulae above correspond to the class of single-valued polylogarithms discussed in many places in the literature. The functions (\ref{QLresult}) correspond to the graphical functions nicely discussed in \cite{Schnetz:2013hqa,Borinsky:2022lds}. However, to our knowledge the functions (\ref{ZLresult}) have not been discussed in terms of graphical functions until now. Below we show that they correspond to conformal ladder graphs of a two-dimensional CFT.

It is useful to introduce the differential operators
\begin{align}
\label{D}
\hat{\bf D}&=\frac{1}{\beta^2}\frac{\partial}{\partial m^2}=\frac{1}{2\ln|z|}(z\partial_z+\bz\partial_\bz)\,,\\
\label{L}
\hat{\,\bf L}&=\frac{i}{\beta}\frac{\partial}{\partial\mu}=(z\partial_z-\bz\partial_\bz)\,.
\end{align}
Explicit calculations yield the following set of first order differential equations \cite{Petkou:2021zhg}
\begin{align}
\label{Oaverage}
&\langle\cO\rangle_L=-\beta\,\hat{\bf D}\,\ln Z_L=\b\a^2\ln Z_{L-1}\\
\label{Qaverage}
&\langle\cQ\rangle_L=\hat{\,\bf L}\,\ln Z_{L}=-\hat{\bf D}\cdot\langle\cQ\rangle_{L+1}/\a^2\,.
\end{align}
Notice that $\hat{\bf D}$ acts on $\ln Z_{L}$ and $\langle\cQ\rangle_L$ as a dimension lowering operator.  Introducing the Laplacian in the variables $m$ and $\mu$ as
\begin{equation}
    \hat{\bf \Delta}=4\b^2\,z\bz\partial_z\partial_{\bz}=\frac{\partial^2}{\partial m^2}+\frac{\partial^2}{\partial\mu^2}\,,
\end{equation} 
we further find  
\begin{equation}
    \label{Laplacian}
    \hat{\bf \Delta} f_L(z,\bz)=-4\beta^2 L \alpha^2f_{L-1}(z,\bz)\,,
\end{equation}
for $f_L(z,\bz)=\{\ln Z_L,\langle\cQ\rangle_{L}\}$.
We can combine (\ref{Laplacian}) with (\ref{Oaverage}), (\ref{Qaverage}) to obtain the second order equation 
\begin{equation}
    \label{LaplacianI}
    \left[m^2\hat{\bf \Delta} -4L\beta^2m^2\hat{\bf D}\right]f_{L}(z,\bz)=0\,.
\end{equation}
Notice that $m^2\hat{\bf \Delta}$ is the Laplacian on the upper half plane  ${\mathbb H}_2$ with coordinates $m,\mu$ and $2\beta^2m^2\hat{\bf D}=m(\partial/\partial m) $ is the radial derivative. Equation (\ref{LaplacianI}) is reminiscent of similar results for partition functions in \cite{Alessio:2021krn} where the a connection to the huge literature of string scattering amplitudes \cite{Vanhove:2018elu,Vanhove:2020qtt,Gerken:2020yii} was noted.  Another interpretation of (\ref{LaplacianI}) is as the Laplace-Beltrami operator of $AdS_{2L+2}$ with metric
\begin{equation}
    \label{AdS}
    ds^2=\frac{1}{m^2}\left(dm^2+d\mu^2+\sum_{i=1}^{2L}dx^idx^i\right)\,,
\end{equation}
acting on functions of just $m$ and $\mu$. Since $m$ and $\mu$ parametrize relevant deformations of a free CFT, such an interpretation may be related to RG flow. 

\emph{Conformal graphs as thermal partition functions}---
We will now show that formulae (\ref{ZLresult}) and (\ref{QLresult}) arise in an apparently unrelated context: as four-point correlators in conformal fishnet models. The latter are particular limits of the generalised bi-scalar theory in $D$-dimensions introduced in \cite{Kazakov:2018qbr} with Lagrangian
\begin{align}
\label{fishnet}
    {\cal L}=&N_c\Tr\left[\phi_1^\dagger(-\partial^2)^{\omega}\phi_1+\phi_2^\dagger(-\partial^2)^{\frac{D-2\omega}{2}}\phi_2\right.\nonumber \\
&\left. \hspace{3cm}+a_{D,\omega}^2\phi_1^\dagger\phi_2^\dagger\phi_1\phi_2\right]\,.
\end{align}
$\phi_{1,2}$ belong to the adjoint of $SU(N_c)$, $\omega\in\left(0,\frac{D}{2}\right)$ and coupling $a_{D,\omega}^2$ is classically dimensionless. We consider the 4pt function 
\begin{equation}
G^{(L)}_{D,\omega}(\{x_i\})=\langle\Tr\left[\phi_2^L(x_1)\phi_1(x_3)\phi_2^{\dagger L}(x_2)\phi_1^{\dagger}(x_4)\right]\rangle\,,
\end{equation}
whose leading $N_c$ contribution comes from a unique $L$-loop conformal ladder graph. It is well known that due to conformal invariance $G^{(L)}_{D,\omega}$ depends on two conformal ratios, or equivalently a complex variable $z$,  and can be represented by an integral of the form depicted in Fig. 1.
\begin{figure}[h]
\centering
\includegraphics[width=0.45\textwidth]{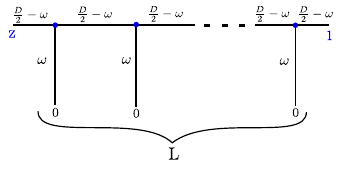}
\caption{The graph contributing to  $G^{(L)}_{D,\omega}$.}
\end{figure}
For $D=4, \omega=1$ the model coincides with the original four-dimensional conformal fishnet CFT introduced in \cite{Gurdogan:2015csr},  and then $G^{(L)}_{4,1}$ is proportional to the Davydychev-Usuykina $L$-loop conformal ladder graphs \cite{Usyukina:1992jd,Usyukina:1993ch}.  Up to overall  normalizations and using (\ref{Oaverage}) we verify that
\begin{equation}
\label{GL41}
\tilde{G}^{(L)}_{4,1}(z,\bz)=\frac{1}{L!}\frac{1}{z-\bz}\langle\cQ\rangle_L(z,\bz)\,,
\end{equation}
when we set $a^2_{4,1}=\a^2$. In writing (\ref{GL41}) we have identified: i) the variable $z$ representing conformal ratios on the l.h.s. with the modular-like parameter $z$ of the thermal QFT on the r.h.s. and ii) the number of loops $L$ on the l.h.s. with $L=(d-1)/2$ on the r.h.s.

For $D=2,\omega=1$ the model (\ref{fishnet}) is singular as $G^{(L)}_{2,1}$ would seem to vanish. \footnote{The $x$-space two-point function with Lagrangian ${\cal L}=\phi(-\partial^2)^a\phi$ in $d=2L+1$-dimensions is $C_\phi^L(a)/x^{2L+1-2a}$, with $C_{\phi}^L(a)=\Gamma(L+1/2-a)/\Gamma(a)4^a\pi^{L+1/2}$.} Nevertheless, a nonzero result can be obtained if we define the effective coupling 
\begin{equation}
    \label{a21}
    \tilde{a}_{D,\omega}=a_{D,\omega}\frac{1}{\Gamma(D/2-\omega)}\,,
\end{equation} 
that remains finite as $D\mapsto 2, \omega\mapsto 1$. Then, following the graph-building techniques introduced in \cite{Derkachov:2018rot,Olivucci:2021cfy,Derkachov:2021ufp,Derkachov:2023xqq} we can show that the appropriately normalised 4pt function of Fig. 1  is given by
\begin{equation}
\label{GL21}
    \tilde{G}^{(L)}_{2,1}(z,\bz)=\tilde{a}_{2,1}^{2L}\sum_{m\in\mathbb{Z}}\int d\nu \frac{(z\Bar{z})^{i\nu}(z/\Bar{z})^{m/2}}{(\frac{m^2}{4}+\nu^2)^{L+1}}\,.
\end{equation}
Since $|z|<1$ we compute the integrals above using contour intergation. When $m\neq 0$ we can close the contour from below and pick up the residues in the lower half complex plane. We obtain
\begin{align}
\label{GL21m}
    &\hspace{-.1cm}\sum_{m\neq 0}\int d\nu \frac{(z\Bar{z})^{i\nu}(z/\Bar{z})^{m/2}}{(\frac{m^2}{4}+\nu^2)^{L+1}}=  \nonumber \\
    &= \frac{2\pi}{L!}\sum_{n=0}^{L} \frac{ (2L-n)! (-2\log |z|)^n}{(L-n)!n!} 2\Re[Li_{2L+1-n}(z)].
\end{align}
For $m=0$ the contour integral appears to be zero, but there is a pole on the real axis. 
Taking the Cauchy principal value we obtain 
\begin{equation}
    -\int_{C_\epsilon} d\nu \frac{|z|^{2i\nu}}{\nu^{2L+2}}=-i\int_\pi^{2\pi} d\theta \frac{\exp{(2i\epsilon\log|z| e^{i\theta})}}{\epsilon^{2L+1}e^{i(2L+1)\theta}}\,.
\end{equation}
For $\epsilon\mapsto 0$ we encounter $2L+1$ divergent terms, which we discard, and a finite contribution which reads
\begin{equation}
\label{GL210}
    -i\int_\pi^{2\pi} d\theta \frac{(2i\log|z|)^{2L+1}}{(2L+1)!}=(-)^L\pi\frac{(2\log|z|)^{2L+1}}{(2L+1)!}\,.
\end{equation}
Putting together (\ref{GL21m}) and (\ref{GL210}) we finally obtain
\begin{equation}
\label{GL21Zl}
    G^{(L)}_{2,1}(z,\bz)=\frac{2\pi}{L!}\ln Z_L(z,\bz)\,,
\end{equation}
when we set $\tilde{a}_{2,1}^2=\alpha^2$. This is one of the main results of the present work. Notice that the leading ``zero temperature" contributions in (\ref{GL21}) and (\ref{ZL}) arise after the subtraction of a finite number of divergent terms. Acting with $\hat{\,\bf L}$ on both sides of (\ref{GL21Zl}) and using  (\ref{Qaverage}) we see that the ladder graphs of the four-dimensional CFT are derivatives of the corresponding ladder graphs of the two-dimensional CFT. This dimension-shift property between conformal ladder graphs generalises to all even dimensions.

\emph{Twisted thermal one-point functions and multiloop conformal graphs}--- 
The thermal one-point functions $\langle\cO\rangle_L$ and $\langle\cQ\rangle_L$ appear in the expansion of the thermal two-point function $\langle\phi^\dagger(x)\phi(0)\rangle=g^{(L)}(\tau,\mathbf{x})$. This motivates us to ask whether thermal one-point functions of higher spin operators are also related to conformal ladder graphs.  It is usually highly nontrivial to calculate thermal one-point functions in a generic QFT. However, for a CFT with a complex scalar $\phi(x)$ having dimension $\Delta_\phi$ in $d=2L+1$ we have \footnote{$x^\mu=(\tau,\mathbf{x})$ are coordinates on the thermal geometry
$S^1_{\beta}\times\mathbb{R}^{d-1}$ with period $\tau\sim\tau+\beta$,
$r=|x|$, $\theta\in[0,\pi]$ is a polar angle in $\mathbb{R}^{d-1}$. $C_{s}^{\nu}(\cos\theta)$ are Gegenbauer
polynomials. }
\begin{equation}
\label{phiphi}
g^{(L)}(\tau,\mathbf{x})=
\sum_{{\cO}_s}a^L_{\cO_s}\left(\frac{r}{\beta}\right)^{\Delta_{\cO_s}}
\frac{C_s^{\nu}(\cos\theta)}{r^{2\Delta_\phi}}\, ,
\end{equation}
where $\nu = d/2-1$. The main assumption behind (\ref{phiphi}) is the existence of a conformal OPE at zero temperature such that $\phi^\dagger\times\phi$ can be expanded in a sum of 
quasiprimary operators ${\cO}_s$ with definite spins $s$ and scaling dimensions
$\Delta_{\cO_s}$. The latter are represented by symmetric, traceless rank-$s$ tensors, and their one-point functions depend on a single parameter which is proportional to the coefficient $a^L_{\cO_s}$. For example, for free massless complex scalars when $\Delta_\phi=L-1/2$ one obtains \cite{Petkou:1998fb,Iliesiu:2018fao,Petkou:2018ynm}
\begin{equation}
\label{azeta}
a^L_{\cO_s}=2C_\phi^L(1)\zeta(2L-1+s)\,,\,s=0,2,4...\,.
\end{equation}
In that case, only symmetric and conserved higher-spin operators with dimensions $\Delta_{\cO_s}=d-2+s$ and even spin $s$ appear in (\ref{phiphi}). Each term in the sum (\ref{phiphi}) is of the form $r^s C_s^{L-1/2}(\cos\theta)$ and we find $\Box_d G^{(L)}(x)=0$ with $\Box_d$ the $d$-dimensional Laplacian. This is the usual free field theory result away from the origin. 

In nontrivial CFTs the operator spectrum, their scaling dimensions and $a^L_{\cO_s}$ change in a way determined by the dynamics, hence the thermal two-point function does not satisfy a simple equation in general, although the form of the expansion (\ref{phiphi}) remains the same. The latter property is not expected to be true in a generic QFT.  
Nevertheless, remarkably, the thermal two-point functions of the complex scalars $\phi(x)$ in the the massive free theory with partition function (\ref{ZL}) does admit an expansion of the form (\ref{phiphi}) and contains a part that is annihilated by the $d$-dimensional Laplacian, albeit with different coefficients $a^L_{\cO_s}$ from (\ref{azeta}). This might not be surprising as the theory is Gaussian, nevertheless the theory is not generically a CFT.
$g^{(L)}(\tau,\mathbf{x})$ is obtained as the Fourier transform of the the unit normalised momentum space two-point function 
with twisted boundary conditions [9] on $S^1_\beta$. 
We obtain (setting $\b=1$ for simplicity)
\begin{equation}
\label{GL}
g^{(L)}(\tau,\mathbf{x})=
\frac{1}{(2\pi)^{\nu}}\sum_{n=-\infty}^\infty\!e^{i\mu n}\left[\frac{m}{|X_n|}\right]^{\nu}\!\!K_{\nu}(m|X_n|)\,,
\end{equation}
with $X_n=(\tau-n,\mathbf{x})$ and $K_\nu$ the modified Bessel functions. 
The coefficients $a^L_{\cO_s}$ can be calculated from (\ref{GL}) using the inversion method of \cite{Iliesiu:2018fao}, as it was done in \cite{Petkou:2018ynm}, but taking now care that the two-point function is complex so that the discontinuities along the cuts in the positive and negative $r$-axis are complex conjugates \footnote{With our normalization the
unit operator $\mathds{1}$ is the unique operator with dimension zero, and
here $a_{\mathds{1}}=C^{L}_{\phi}(1)$.}. We focus on the part of $G^{(L)}(\tau,\vec{x})$ that is annihilated by the $d$-dimensional Laplacian, namely to the contribution of the would-be higher-spin currents with dimensions \mbox{$\Delta_{\cO_s}=2L-1+s$}. We obtain
\begin{eqnarray}
    \label{aOs}
    \hspace{-1cm}a^{L}_{\cO_s}&=&\frac{\Gamma\left(L-\frac{1}{2}\right)}{\Gamma\left(L+s-\frac{1}{2}\right)(4\pi)^L 2^{2s}}\nonumber \\
    &\times&\sum_{n=0}^{L-1+s}\frac{2^n}{n!}\frac{(\beta m)^n(2L-2+s-n)!}{(L-1+s-n)!}\nonumber \\
    &\times&\left[Li_{2L-1+s-n}(z)+(-1)^s Li_{2L-1+s-n}(\bar{z})\right]\,.
\end{eqnarray}
If the theory were a CFT we would associate the coefficients $a^L_{\cO_s}$ with thermal one-point functions of conformal quasiprimary operators. 
For generic values of $m$ and $\mu$ this is more complicated. For example, $a^L_{\cO_2}$ represents the  contribution of a rank-2 symmetric traceless tensor which is {\it not} the the energy momentum tensor of the massive theory since the latter has nonzero trace.   Nevertheless, the coefficients $a^L_{\cO_0}$ and $a^L_{\cO_1}$ {\it do} represent the thermal one-point functions of the operators $\cO$ and $\cQ$ as they have been independently calculated in  (\ref{Oaverage}), (\ref{QLresult}). Explicitly we have
\begin{equation}
    \label{aOaQ}
    a^L_{\cO_0}=\frac{1}{(4\pi)^L\beta\alpha^{2L}}\langle\cO\rangle_L\,,\,\,a^L_{\cO_1}=\frac{1}{(4\pi)^L\alpha^{2L}}\frac{1}{2}\langle\cQ\rangle_L\,.
\end{equation}
Using (\ref{Oaverage}), (\ref{Qaverage}) we see that for $z=\bz=1$ the above  reduce to (\ref{azeta}) as they should.  
The novel result is that {\it all} coefficients $a^L_{\cO_s}$ with $s\geq 2$ are related to $L$-loop conformal graphs by virtue of the following recursion relations shown by brute force calculations
\begin{equation}
    \label{recursion}
    a^L_{\cO_{s+2}}=\frac{2\pi}{2L-1}a^{L+1}_{\cO_s}+\frac{(m\beta)^2}{(2L-1+2s)(2L+1+2s)}a^L_{\cO_s}
\end{equation}
Consequently, the part of the twisted thermal two-point function (\ref{GL}) that  is annihilated by the $d$-dimensional Laplacian is a generating function for (linear combinations) of $L$-loop conformal ladder graphs.

Our (\ref{recursion}) implies that we can associate a ``spin" to a certain combination of $L$ and $L-1$-loop conformal ladder graphs. This is evident for $s=0$, and it can be generalised for all $s$. We do not yet have an understanding of this ``spin" from the point of view of the conformal graphs, but from the thermal field theory point of view it can be given a physical interpretation in terms of the underlying free field theory dynamics.
However, we believe that they  have a simpler underlying physical interpretation. For example, (\ref{recursion}) corresponds to a standard thermodynamics relationship for $s=2$. To see that note that from the twisted partition function $Z_L$ with Hamiltonian of the form $H=H_0+m^2\cO+i\mu\cQ$ one can derive the following general result
\begin{equation}
    \label{thermodynamics}
    \langle H\rangle_L=\frac{d-1}{\beta}\ln Z_L+2 m^2\langle\cO\rangle_L+i\mu\langle\cQ\rangle_L\,,
\end{equation}
where $\langle H\rangle_L=-\langle t_{\tau\tau}\rangle_L$ with  $t_{\m\n}$ the energy momentum tensor of theory. For nonzero $m$ and $\mu$ this is not traceless, but for the massless free complex scalar with imaginary chemical potential we can construct a traceless spin-2 operator ${\cal T}_{\m\n}$ with ${\cal T}_{\tau\tau}=t_{\tau\tau}+2m^2\cO/d+i\mu\cQ$. Then (\ref{thermodynamics}) becomes
\begin{equation}
    \label{thermodynamics2}
    -\langle{\cal T}_{\tau\tau}\rangle_L=\frac{d-1}{\beta}\ln Z_L+2 m^2\frac{d-1}{d}\langle\cO\rangle_L\,.
\end{equation}
The general relation connecting $a^L_{\cO_2}$ with the ${\cal T}_{\tau\tau}$ is \footnote{We use the standard free CFT results for the three-point function coupling $g_{\phi^\dagger\phi {\cal T}}$, the normalization of the two-point function of ${\cal T}_{\m\n}$, $C_{\cal T}$ and we take into account that a complex scalar corresponds to two real scalars. $S_L=2\pi^{L+1/2}/\Gamma(L+1/2)$ is the surface of the $2L+1$-dimensional unit sphere.}
\begin{eqnarray}
    \label{aLO2}
    \frac{(4\pi\a^2)^L}{\beta}a^L_{\cO_2}&=&\frac{2g_{\phi^{\dagger}\phi T}}{(d-1)(d-2)C_{T}}\langle T_{00}\rangle_L\nonumber \\
    &=& -\frac{C^L_{\phi}(1)S_L}{2(d-1)}\langle{\cal T}_{\tau\tau}\rangle\,.
\end{eqnarray}
Using then (\ref{aOaQ}), (\ref{Oaverage}) and (\ref{Qaverage}) we can verify that (\ref{thermodynamics2}) coincides with (\ref{recursion}). We believe that similar arguments relating tracefull and traceless higher spin operators of the  massive free scalar theory can provide a physical understanding for  (\ref{recursion}) for general $s$.

\bigskip
\emph{Discussion and outlook}---  In this work we have connected two seemingly unrelated quantities: twisted partition functions of a massive free complex scalar field in $d=2L+1$ dimensions,  and four-point conformal $L$-loop ladder graphs. The reason for such a relationship is that they both satisfy the same sets of differential equations. For the partition functions these are given by (\ref{LaplacianI}). For the conformal ladder graphs they are the differential equations discussed in number of earlier works on conformal integrals (i.e. eq. 2.15 in \cite{Drummond:2012bg}). This common property begs for a deeper explanation.  

Our results draw a unifying picture for the thermal one-point functions $a^L_{\cO_s}$ in massive free complex scalar theories. This is depicted in Fig. 2. By the algebraic relations (\ref{recursion}) they are all ultimately given by  $a^L_{\cO_0}$ or $a^L_{\cO_1}$, and then by the action of the differential operators $\hat{\bf D}$ and $\hat{\,\bf L}$ to the $\langle\cO\rangle_0$ and $\langle\cQ\rangle_0$ of the harmonic oscillator model (\ref{Z0def}). 
\begin{figure}[h]
\centering
\includegraphics[width=0.4\textwidth]{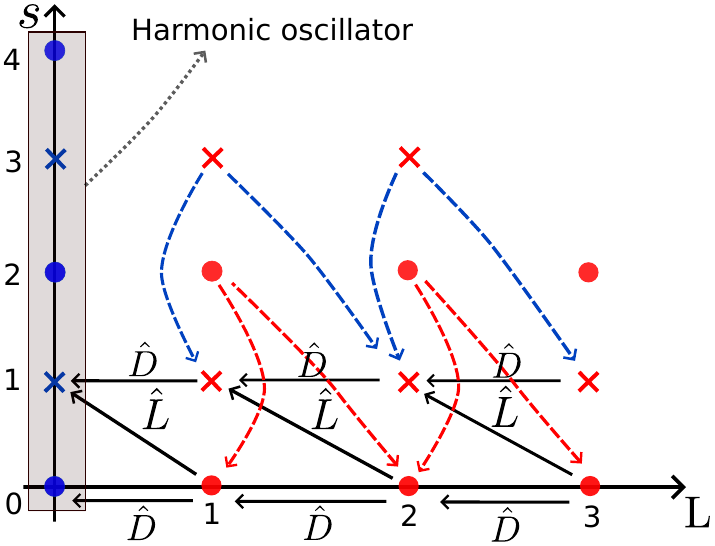}
\caption{Differential (solid lines) and algebraic (dashed lines) relationships among the $a^L_{\cO_s}$.}
\end{figure}

There are many questions that arise from our observations. It would be interesting to understand the possible relationship of our results to the integrability of fishnet models. It would also be interesting to connect our results to works that relate partition functions and string amplitudes. Another question would be to connect our approach to studies of non-integrable deformations of thermal CFTs (i.e. see \cite{David:2023uya, Diatlyk:2023msc,Benedetti:2023pbt} for interesting recent works).

We close with some remarks. Our iterated integral formula (\ref{ZL}) when applied to $\langle\cQ\rangle_L$ gives for $L=1$
\begin{equation}
    \label{Q1}
    \langle\cQ\rangle_1=(-2\a^2)\int_0^{|z|}\frac{d|z'|}{|z'|}\ln|z'|\langle\cQ\rangle_0=4i\a^2D(z)\,,
\end{equation}
where $D(z)=\Im\left[Li_2(z)+\ln|z|\ln(1-z)\right]$ is the celebrated Bloch-Wigner function that gives the volume of an ideal tetrahedron in three-dimensional hyperbolic space ${\mathbb H}^3$ with vertices in $\partial{\mathbb H}^3$ \cite{Zagier2}. It is then amusing to note that $\langle\cQ\rangle_0$ itself has a geometric interpretation. Indeed, 
\begin{equation}
    \label{triangle}
    \langle\cQ\rangle_0=\frac{z-\bz}{(1-z)(1-\bz)}\,.
\end{equation}
and setting $z=e^{i\phi}(b/a)$ with $\cos\phi=(a^2+b^2-1)/2ab$ we find that $\langle\cQ\rangle_0/4i=\frac{1}{2}ab\sin\phi$ gives the area of a triangle whose side lengths are $a,b$ and 1, and $\phi$ the angle between $a$ and $b$. Then (\ref{Q1}) gives the volume of an ideal hyperbolic tetrahedron as an integral of the area of a triangle. One then wonders if there is a geometric interpretation for the higher order iterated integrals in (\ref{ZL}). We should further note that  $\langle\cO\rangle_0$ also has an interpretation as an area, but we are not aware of a nice geometric interpretation of $\langle\cO\rangle_1$. 

Another  observation is that
   $\langle\cQ\rangle_0=-8\pi\rho^{D=4}_{m_0\mapsto m_1+m_2}$
where $\rho^D_{m_0\mapsto m_1+m_2}$ is the $D$-dimensional  $1\mapsto 2$ decay phase space of relativistic massive particles. Since 
\begin{equation}
\label{Kallen}
\rho^{D=4}_{m_0\mapsto m_1+m_2}=\frac{1}{8\pi}\sqrt{\lambda (1,m_1^2/m_0^2,m_2^2/m_0^2)}
\end{equation}
with $\lambda(a,b,c)=a^2+b^2+c^2-2ab-2ac-2bc$ the K\"all\'en triangle function, we see that if we set \mbox{$a=m_1/m_0$} and \mbox{$b=m_2/m_0$}, then $\langle\cQ\rangle_0$ represents the phase space for a virtual process with $\lambda <0$ \cite{Kaldamae:2014fua}. Then, our (\ref{ZL}) is reminiscent to eq. (7) of \cite{Delbourgo:2003zi} that gives a recurrent relationship for higher dimensional  $1\mapsto 2$ relativistic phase spaces.

We further note that equations such as (\ref{Oaverage}) and (\ref{Qaverage}) lead naturally to the resummation of infinite series. For example, by virtue of (\ref{Oaverage}) the infinite product  $Z=\prod_{n=0}^{\infty}Z_n$ satisfies the inhomogenous first order equation
\begin{equation}
    \label{Z}
    (\hat{\bf D}+\a^2)\ln Z=-\frac{1}{\beta}\langle \cO\rangle_0\,.     
\end{equation}
This can be integrated to 
\begin{equation}
    \label{resumm}
    \ln Z=-\b e^{-\b^2\a^2m^2}\int^{m^2}\!\!\!e^{\b^2\a^2\tilde{m}^2}\langle\cO\rangle_0\,d\tilde{m}^2\,.
\end{equation}
An analogous result can be derived for the \mbox{$\langle\cQ\rangle=\sum_{n=0}^\infty\langle\cQ\rangle_n$}. Given (\ref{GL41}) and (\ref{GL21Zl}) these are all-loop Borel summations of conformal ladder graphs \footnote{We thank C. Bachas and D. Benedetti for pointing out this to us}. See Appendix A for some additional observations.

Finally, we point out the work \cite{Sinha:2022sdo} where 2-2 scattering amplitudes are given in terms of a dispersive integral over generating functions of knot polynomials (see e.g. (12) and (23) of that reference) \footnote{We thank the referee for bringing this work to our attention.}. The latter generating functions written in terms of the variables $z,\bar{z}$ correspond to thermal averages of certain bilinear operators in a q-deformed harmonic oscillator, much like our $\langle O\rangle_L$ and $\langle Q\rangle_L$. We find the connection of the approach in \cite{Sinha:2022sdo} and our results quite intriguing and we believe that in deserves further study.

\medskip
\emph{Acknowledgements}--- A.C.P. has benefited over an extended period of time from  discussions and correspondence with G. Barnich, A. Mcleod, O. Schnetz, J. Usovitsch, and P. Vanhove. M.K. and A. C. P. would like to thank G. Katsianis for useful discussions and collaboration at the early stages of this work. 
We wish to thank E. Olivucci for patiently explaining his work to us. We are grateful to the Mainz Institue for Theoretical Physics (MITP) of the Cluster of Excellence PRISMA$^{+}$ (Project ID 39083149), for its hospitality and its partial support during the workshop on ``Thermalization in Field Theories" in July 2023 where part of this work was completed. The work of S. L. , A. C. P.  and M. V. was supported by the Hellenic Foundation for Research and Innovation (H.F.R.I.) under the \textsl{First Call for H.F.R.I. Research Projects to support Faculty members and Researchers and the procurement of high-cost research equipment grant} (MIS 1524, Project Number: 96048).

\medskip
\emph{Appendix A. } --- \emph{Futher observations} -- Applying $\hat{\,\bf L}$ to (\ref{resumm}) gives
\be\label{BDsum}
(\hat{\bf D}+\alpha^2)\langle Q\rangle=\hat{\bf D}\langle Q\rangle_0\,
\ee
where $\langle Q\rangle=\sum_{L=0}^\infty\langle Q\rangle_L$.  By virtue of (\ref{GL41}) this sum can be Borel transformed into the Broadhurst-Davydychev infinite sum of the $L$-loop conformal ladder graphs in four-dimensions \cite{Broadhurst:2010ds}, see also \cite{Giombi:2020enj,Caetano:2023zwe}. Indeed the solution of the first order equation (\ref{BDsum}) is
\be\label{BDsum1}
\langle Q\rangle=\beta^2 e^{-\beta^2\a^2m^2}\int^{m^2}e^{\beta^2\alpha^2\tilde{m}^2}\hat{\bf{D}}\langle Q\rangle_0d\tilde{m}^2\,,
\ee
and can be thought of as a series of the form \mbox{$\langle Q\rangle \equiv A(z)=\sum_{n=0}^\infty a_n z^n$} with $z=\a^2$. 
Its Borel transform series ${\cal B}\left[A\right](t):= \sum_{k=0}^{\infty}\frac{a_{k}}{k!}t^{k}$ is given by the contour integral 

\beq
\label{Borel_transform_inverse}
{\cal B}\left[A\right](t)= \frac{1}{2\pi i} \int_{C}\frac{dz}{z} e^{z} A(t/z)
\eeq
where ${C}$ is the Hankel contour \footnote{We use the convention that C starts at $\infty-i \epsilon$ with $\epsilon>0$, then encircles (0,0) counterclockwise and ends up to $-\infty+ i \epsilon$.}. 
Using the following integral representation of the Bessel function
\beq
\label{Bessel_Hankel_integral_2}
J_{\nu}(z)=\frac{(\frac{1}{2}z)^{\nu}}{2\pi i}\int_{C}dt \frac{1}{t^{\nu +1}}e^{t- \frac{z^2}{4t}}\,,
\eeq
we obtain
\beq
\label{BDeq2}
\begin{split}
{\cal B}[Q](t)&= 
\beta^2 \frac{1}{2\pi i}\int^{m^2}\!\!\!\!\!\!d\tilde m^2(\hat{\bf D}\langle Q\rangle_0)\int_{C}\frac{du}{u}e^{u - \frac{t\beta^2}{u}(m^2- \tilde m^2)}
\\&
=\beta^2 \int^{m^2}\!\!\!\!\!\!d\tilde m^2 J_{0}(2\beta\sqrt{t(m^2 -\tilde m^2)})(\hat{\bf D}\langle Q\rangle_0)
\end{split}
\eeq
Using then 
\beq
\label{D_Q_0}
\hat{\bf D}\langle Q\rangle_0= \frac{i}{2\beta m}\frac{\text{sinh}(\beta m)\text{sin}(\beta\mu)}{(\text{cosh}(\beta m)-\text{cos}(\beta\mu))^2}\,,
\eeq
and setting  $t=-\frac{\kappa^2}{4}$,  $\beta\tilde m=\eta$, $\ell=2\beta m$ and putting the lower bound of the integral to be $+\infty$, (\ref{BDeq2}) coincides with eq. 15 of \cite{Broadhurst:2010ds}. 

\emph{Appendix B. } --- \emph{The relativistic thermal gas} --- 

The one-particle density of states $\rho_L(\omega;m)$ for the relativistic thermal gas in $d=2L+1$-dimensions 
is found as usual by considering the system in a $(d-1)$-dimensional spatial cubic box of volume $V_{d-1}=\ell^{d-1}$ with quantized  momentum  $\vec{p}=\left(\frac{2\pi}{\ell}n_1,...,\frac{2\pi}{\ell}n_{d-1}\right)=\frac{2\pi}{\ell}\vec{n}
$.
The number of modes having momenta inside the spherical shell bounded by $|\vec{p}|$ and $|\vec{p}|+d|\vec{p}|$ in $d=2L+1$ dimensions is 
\be\label{dn}
dn=\left(\frac{\ell^2}{4\pi^2}\right)^L |\vec{p}|^{2L-1}d|\vec{p}|\int d\Omega_{2L}\,,
\ee
with $\int d\Omega_{2L}=2\pi^L/\Gamma(L)$. Using then the dispersion relation $\omega^2=\vec{p}\,^2+m^2$, for $\rho_L(\omega;m) \equiv dn/d\omega$ we obtain 
\begin{equation}
    \label{eq: rho_L}
    \rho_L(\omega;m) = \frac{2\a^2\b^2}{(L-1)!}\omega (\omega^2-m^2)^{L-1}\,,
\end{equation}
which when substituted in (\ref{Zd}) gives 
\be\label{Zdexpl}
\ln Z_L=\frac{2\a^2\b^2}{(L-1)!}\int_m^\infty\omega d\omega(\omega^2-m^2)^{L-1}\ln Z_0\,,
\ee
or alternatively (\ref{ZL}) in terms of the real variable \mbox{$\omega$ $|z|=e^{-\beta m}$}. 
We can now apply our differential operators $\hat{\bf D}$ and $\hat{\,\bf L}$ to this and obtain the integral representations of all our thermal one-point functions. In particular, applying $\hat{\,\bf L}$ to (\ref{Zdexpl}) we will get the integral representation of the $L$-loop conformal ladder graphs given in eq. (2.20) of \cite{Basso:2021omx} - for purely imaginary $\phi$ and up to and overall $1/L!$.

\end{document}